# Rapid contemporary evolution and clonal food web dynamics

Laura E. Jones (1), Lutz Becks (1, 2), Stephen P. Ellner (1), Nelson G. Hairston Jr. (1)
Takehito Yoshida (3), and Gregor F. Fussmann (4)

(1) Department of Ecology and Evolutionary Biology, Corson Hall, Cornell University,
Ithaca NY 14853-2701 USA.
(2) Present address:  Department of Ecology and Evolutionary Biology, 25 Willcocks St.,
Toronto, Ontario, M5S 3B2, Canada.
(3) Department of General Systems Studies, University of Tokyo, 3-8-1 Komaba,
Meguro, Tokyo 153-8902, Japan.
(4) Department of Biology, McGill University, 1205 ave. Docteur-Penfield, Montreal,
Quebec, H3A 1B1, Canada.

## Abstract

Character evolution that affects ecological community interactions often occurs
contemporaneously with temporal changes in population size, potentially altering the
very nature of those dynamics.  Such eco-evolutionary processes may be most readily
explored in systems with short generations and simple genetics. Asexual and cyclically
parthenogenetic organisms such as microalgae, cladocerans, and rotifers, which
frequently dominate freshwater plankton communities, meet these requirements. Multiple
clonal lines can coexist within each species over extended periods, until either fixation
occurs or a sexual phase reshuffles the genetic material. When clones differ in traits
affecting interspecific interactions, within-species clonal dynamics can have major effects
on the population dynamics. We first consider a simple predator-prey system with two
prey genotypes, parameterized with data on a well-studied experimental system, and
explore how the extent of differences in defense against predation within the prey
population determine dynamic stability versus instability of the system. We then explore
how increased potential for evolution affects the community dynamics in a more general
community model with multiple predator and multiple prey genotypes. These examples
illustrate how microevolutionary "details" that enhance or limit the potential for heritable
phenotypic change can have significant effects on contemporaneous community-level
dynamics and the persistence and coexistence of species.

## Introduction

The last three decades have seen an accumulation of studies demonstrating that evolutionary change in ecologically important organismal traits often takes place at the same time and pace as ecological dynamics ("eco-evolutionary dynamics"; Fussmann et al. 2007; Kinnison & Hairston 2007). Species as diverse as single-celled algae, annual plants, birds, fishes, crustaceans, insects, and sheep are found to undergo rapid contemporary evolutionary changes in traits that adapt them within a few generations to changing or new environments (Reznick et al. 1997, Thompson 1998, Hairston et al. 1999, Sinervo et al. 2000, Cousyn et al. 2001, Reznick & Ghalambor 2001, Grant & Grant 2002, Heath et al. 2003, Yoshida et al. 2003, Olsen et al. 2004, Pelletier et al. 2007, Swain et al. 2007). Thus, the old assumption that evolutionary change is negligible on the time scale of ecological interactions is now demonstrably incorrect.

Ultimately, evolution has the potential to transform how we think about and manage ecological systems, with implications for resource management (Heath et al. 2003, Swain et al. 2007, Strauss et al. 2008), conservation of biodiversity (Ferriere & Couvert 2004, Kinnison & Hairston 2007), invasive species management (Lavergne & Molofsky 2007), and ecosystem responses to environmental change (Hairston et al. 2005, Strauss et al. 2008). However, virtually all of the ecological theory developed to address these practical issues invokes, tacitly or explicitly, a separation of time scales between ecological and evolutionary dynamics (see e.g. Ferriere & Couvert 2004). Frequently that's valid, of course, but the accumulating evidence suggests that important questions may be answered incorrectly if we fail to account for the fact that we are studying and managing a "moving target".

Here we use the compound term "rapid contemporary evolution" to refer to heritable changes within a population that occur at a rate fast enough to affect interspecific interactions while they are taking place. The descriptor "rapid", by itself, leaves ambiguous the frame over which evolutionary rates are compared (rapid species diversification on a geological time scale can be quite slow compared with the rate of heritable changes within a population). The adjective "contemporary" (Hendry and Kinnison 1999) identifies the time scale that interests us, but permits the degree of change to be either negligible or dramatic. Our intent is to explore the effect of evolution that both occurs within a few generations (the time scale of ecological dynamics) and is substantial (ecological interaction strength is importantly altered); that is, evolution that is both contemporary and rapid.

Post and Palkovacs (this volume) note that a system of eco-evolutionary feedbacks requires both that phenotypes must affect the environment in which they live and the environment must select on the distribution of genetically based phenotypes. A major component of ecological research has been the identification and measurement of interaction strengths among species, and between species and their chemical and physical environments (e.g., Cain et al. 2008, Molles 2008), and Post and Palkovacs (this volume) provide a table of fine examples. At the same time, there is a growing literature showing that heritable phenotypic differences among individuals have substantial importance for how communities and ecosystems function (e.g., Johnson et al. 2006, Whitham et al. 2006, Urban et al. 2008, Palkovacs et al. this volume). Finally, extensive reviews of the literature (Hendry and Kinnison 1999, Palumbi 2001, Kinnison and Hendry 2001) leave no doubt that selection driven by strong ecological interactions results in marked adaptive evolutionary change that frequently takes place within a few generations.

A challenge, then, is to draw together each of these components within a single system to determine the importance of rapid contemporary evolution to ecological dynamics. Hairston et al. (2005) proposed an approach, first conceived by Monica Geber, in which the relative importance of temporal changes in a feature of the environment and heritable changes within a population are assessed in terms of the magnitudes their contributions to ecological dynamics. They applied it to data on Galapagos finches (Grant and Grant 2002), freshwater copepods (Ellner et al. 1999) and laboratory microcosms (Yoshida et al. 2003). Pelletier et al (2007) applied a similar method to Soay sheep, and Ezard et al. (this volume) use this approach on five populations of ungulates. In each case evolutionary contributions to ecological dynamics were found to be substantial.

Some of the most striking examples of temporal ecological dynamics in nature are found in consumer-resource interactions. Theory indicates and empirical data confirm that population abundances frequently oscillate substantially as interactions vary in strength, and in some cases sign, over time. Many of the clearest examples of rapid contemporary evolution have been documented in these systems including predatory fish and their prey (Hairston and Walton 1986, Reznick et al. 1997, Ellner et al. 1999, Hairston and De Meester 2008), herbivorous insects and the plants they consume (Johnson et al. 2006, Johnson et al. this volume), and parasites and pathogens and their hosts (Dube et al. 2002, Decastecker et al. 2007, Duffy et al. 2007, Eldred et al. 2008). Because of the propensity to oscillate, consumer-resource interactions are particularly good model systems for detailed laboratory study of eco-evolutionary dynamics over both long (Lenski and Travisano 1994, Borras et al. 1998) and short term (Meyer and Kassen 2007). Investigations using simple laboratory microcosms of real consumer and resource species combined with mathematical models that incorporate observed interactions and predict resulting dynamics have proven to be particularly fruitful for uncovering some of the potential and diversity of eco-evolutionary dynamics (e.g., Yoshida et al. 2003, Meyer et al. 2006, Jones and Ellner 2007, Yoshida et al. 2007).

In this paper we explore how evolutionary dynamics in traits that influence the strength and nature of interspecific interactions affect the temporal dynamics of species, and the reciprocal effects of population dynamics on maintenance of the heritable variation that allows evolution to continue. We begin by reviewing some of our relevant prior experimental and theoretical work on rapid contemporary evolution in predator-prey systems, and then use extensions of our previous models to address the following general questions:

1. How does evolution affect the dynamics of ecological communities, and their persistence, robustness and stability?

2. What are the dynamic possibilities when a wide range of heritable trait variation is present in a system, and what are the consequences of diminished genetic diversity and the correspondingly reduced potential for evolutionary change?

Theory linking evolution and ecology has been published for decades, but theory unfettered by data can predict (and has predicted) any conceivable ecological dynamics, including stability, instability, chaos, and criticality (the boundary between two different types of dynamics, e.g., stability versus cycles). To make specific predictions possible, our theory is closely tied to our experimental work on predator-prey chemostats that are laboratory analogs of freshwater planktonic food webs.

Natural freshwater plankton systems are often dominated by obligately asexual or facultatively sexual species, so our models assume that trait evolution occurs through changes in the frequency of different clonal lineages. However, models for quantitative trait dynamics in sexual species can be very similar to our models for clone-frequency dynamics, and can even be mathematically identical (depending on assumptions about the genetic variance for the quantitative trait; see Abrams and Matsuda 1997, Abrams 2005).

As a result, our conclusions might also be relevant to sexually reproducing species. Models for clone-frequency change are also mathematically equivalent to models for multispecies interactions, but some predictions about clonal evolution result from constraints on the differences among clones within a species, that typically would not be realistic if we were modeling completely different species (Jones and Ellner 2007).

## Background: Predator and Prey in the Chemostat

In laboratory microcosms containing a rotifer, *Brachionus calyciflorus*, consuming an alga, *Chlorella vulgaris*, we observed predator-prey cycles with the classic quarter-period phase lag between the peaks in prey and predator densities when the algal prey were genetically homogeneous. In contrast, with a genetically variable prey population, cycle period increased 3-6 fold, and predator and prey cycles were exactly out of phase (Shertzer et al. 2002, Yoshida et al. 2003; Fig. 1), a phenomenon that cannot be explained by conventional predator-prey models. The algae in our microcosms vary genetically along a tradeoff curve between defense against rotifer predation and ability to compete for limiting nutrients (Yoshida et al. 2004). As a result, they have an evolutionary response to fluctuations in rotifer density and nutrient availability that qualitatively changes the predator-prey dynamics.

To understand our experimental results, we formulated and analyzed a general model for a predator-prey system with genetic variability in an asexually reproducing prey species (Yoshida et al. 2003, Jones & Ellner 2004, Jones & Ellner 2007, Yoshida et al. 2007). Our results show that the qualitative properties of evolution-driven cycles in our experimental system are a general consequence of rapid prey evolution in a predator-prey food chain, occurring in a specific but biologically relevant region of parameter space. Specifically, genotypes with better defense against predation become more common when predators are abundant, and less common (due to poorer competitive ability) when predators are rare. We then confirmed this theoretical prediction experimentally, with close qualitative and quantitative agreement between experimental results and a refined version of our evolutionary model (Yoshida et al. 2003; Fig. 1A-D). Finally, we verified the postulated tradeoff between defense against predation and ability to compete for scarce nutrients (Yoshida et al. 2004).

A basic model for our experimental system is given by equation (1) below; this model will be the starting point for the new theoretical results that we present in this paper:

$$\frac{dS}{dt} = 1 - S - \sum_{i=1}^{2} \frac{x_i mS}{k_i + S}$$

$$\frac{dx_i}{dt} = x_i \left[ \frac{mS}{k_i + S} - \frac{p_i yG}{k_b + Q} - 1 \right] \qquad (1)$$

$$\frac{dy}{dt} = y \left[ \frac{GQ}{k_b + Q} - 1 \right]$$

where $x_1, x_2, y$ are the abundances of the two prey clones and the predator, $S$ is the amount of a limiting substrate required for algal growth, and $Q = p_1 x_1 + p_2 x_2$ is total prey "quality" as perceived by the predator. Time has been rescaled so that the dilution rate (the fraction of medium replaced each day) is 1, and state variables have been rescaled so that the rate of substrate supply is 1 and a unit of prey consumption yields one net predator birth.

The prey types are assumed to be two genotypes within a single species, differing in two ways that we have documented in our experimental systems: their "palatability" $p_i$ which determines their relative risk of being attacked and consumed by the predator, and their ability to compete for scarce resources, represented by $k_i$. There is a tradeoff between defense against predation and the ability to compete for nutrients when nutrient concentration is low (Yoshida et al. 2004), so the genotype with the lower value of $p$ has the higher value of $k$. The models fitted to our experimental data also include predator age-structure (Fussmann et al. 2000, Yoshida et al. 2003), mortality within the chemostat, and a predator functional response that becomes linear (type-I) at low food density, but these features have no qualitative effect apart from stabilizing the system at very low dilution rates (Jones & Ellner 2007). The model's main qualitative predictions about potential effects of rapid contemporary prey evolution on predator-prey cycles are also robust to changes in the functional forms of the prey and predator functional responses (Jones & Ellner 2007). However, it is important for our results that the model does not allow the predator to preferentially attack and consume the less defended prey type. Prey genotypes are attacked in proportion to their abundance, and better defense means a higher chance of surviving an attack. These assumptions have been verified experimentally for our chemostat system (Meyer et al. 2006).

By mathematical and computational analysis of a general version of model (1) we have shown (Jones and Ellner 2007) that evolution-driven cycles occur when defense against predation is effective but cheap: $p_1$ is much smaller than $p_2$, but $k_1$ is not much larger than $k_2$. When the cost of defense is extremely low a surprising phenomenon occurs that we have called "cryptic dynamics": the predator cycles in abundance, but the total prey population remains effectively constant through time (Fig. 1E). We know that the cost of defense is quite low for some genotypes in our algal study species (Meyer et al. 2006), and the cryptic dynamics predicted by the model have been observed in our experimental system (Figure 1F, Yoshida et al. 2007). The constancy of total prey abundance occurs in the model due to near-exact compensation between the defended and undefended genotypes. Each genotype is undergoing large changes in abundance, but as the cost of defense is made very low, the oscillations of the two genotypes become almost exactly out of phase with each other, leading to near-constancy of their sum (Jones and Ellner 2007). In addition, the coefficient of variation in the trait becomes smaller and smaller relative to the coefficient of variation in the predators (compare the dotted curves in Fig 2E versus Fig 2C): only the predator continues to have cycles whose peaks are much higher than the troughs. Since defense is demonstrated to be cheap, or even free, in a number of natural and laboratory systems, especially those involving organisms with short generation times (Andersson & Levin 1999, Yoshida et al. 2004, Gagneux et al. 2006), cryptic dynamics may be more common than hitherto anticipated.

## Results I: dynamic effects of prey genetic variability

The phenomenon of cryptic dynamics shows that the dynamics of a predator-prey interaction are not just determined by the presence or absence of contemporary evolution. The details matter, and the nature and extent of genetic variability for traits affecting the interaction are important details.

Our goal in this paper is to explore theoretically how changes in the suite of genotypes that are present in the prey and predator populations can affect dynamics at the population and community levels. We begin, in this section, by considering genetic variation in prey alone, with prey clones arrayed along a tradeoff curve between competitive ability and defense against predation. Given a tradeoff curve, how is the qualitative nature of the population-level dynamics affected by the presence or absence of specific genotypes along the curve, such as the presence or absence of well-defended types, or the number of alternative clones filling in the range of variation from low to high levels of defense? And how does the shape of the tradeoff curve affect these predictions?

**Systems with two extreme prey types**

A frequent outcome of competition between two prey clones in our model is selection for two extreme types: either the very least and most defended among the clones present in the population, or clones very near to those extremes. In such cases, the range of variation present (the values of $p_1$ and $p_2$) determines the outcome, and the shape of the tradeoff curve is irrelevant.

Figure 2 summarizes how the population dynamics are affected by varying the range of palatability when only two extreme clones are present. When the better-defended clone has very effective defense ($p_1 \leq 0.1$), theory predicts the type of evolutionary cycles (EC in Fig. 2) observed by Yoshida et al. (2003), or possibly cryptic cycles (Yoshida et al. 2007), depending on the cost of defense. When defense is less effective, the system goes to a steady state (SS in Fig. 2). For a narrow range of $p_1$ values (e.g., $p_1 \approx 0.14$ in Fig. 2), both prey types coexist at steady state: the fraction of the defended prey clone is high but less than 100%. This is followed by a substantial range of defense levels where the steady state involves only the defended clone and the predator. Eventually, for moderate to low levels of defense ($0.3 \geq p_1 < p_2$ in Fig. 2), there are classical consumer-resource cycles (CRC in Fig. 2) with the predator and only the defended clone. What happens when defense is nearly ineffective ($p_1 \approx p_2$) is very sensitive to the specifics of the tradeoff curve. However the qualitative pattern shown in Fig. 2 for highly to moderately effective defense is robust to changes in parameter values, and also to some qualitative changes in the model (i.e., the form of the predator functional response, inclusion versus omission of predator age structure (Jones & Ellner 2004, 2007)).

The pattern predicted in Fig. 2 is somewhat counter-intuitive, because it says that improved defense by a defended clone is beneficial to the undefended clone. Specifically, an undefended clone can persist in the system only when the defended clone is nearly invulnerable to predation. What causes this pattern is indirect facilitation, a reversal of the familiar "apparent competition" scenario. Increased defense by the defended prey (on its own) is bad for the predator, so it is good for undefended prey. As a prey lineage evolves better and better defense, eventually driving the predator population to low numbers, it opens the way for a superior competitor that isn't paying a cost (however small) for anti-predator defense.

Some preliminary results (Fig. 3; L. Becks, *unpublished data*) from chemostat experiments with *Chlamydomonas reinhardii* as the algal prey exhibit the pattern predicted by Fig. 2 with regard to evolutionary cycles versus steady-state coexistence The *Chlamydomonas* system has the advantage that one prey defense trait, namely the formation of clumps that are less easily consumed by the predator, is visible and easily quantified, so we can track the dynamics of this trait along with the population dynamics. We have shown that prey clumping reduces the ability of predators to consume them, and that the tendency to clump is heritable (L. Becks, *unpublished data*). However, clumping may not be the only defense trait that these prey can evolve, so the degree of clumping may not be a complete indication of the level of prey defense. A direct indicator of the effectiveness of prey defense is the relationship between prey abundance and predator population growth. The better defended the prey are, the lower the predator per-capita rate of increase at any given prey abundance, and the higher the prey abundance required for the predator population to increase rather than decrease.

In the experiment shown Fig. 3A, the system settled to a steady state with defended prey (as indicated by the occurrence of clumping). The level of prey defense in this experiment was such that the predator population had per-capita rate of increase $r \approx 0$ at algal density of $\approx 10^5 / \text{ml}$ (days 45-58). The experiment shown in Fig. 3B was run under the same conditions, but was initialized with prey drawn from a population that had been exposed to predation for several months. The prey in this latter experiment developed far more effective defense than in the former experiment, as seen by the fact that (for example) at days 25 and 60 the predator population was decreasing even though the algal density was approximately $2 \times 10^5 / ml$. As predicted in Fig. 2, stronger prey defense was accompanied by a shift from steady state to evolutionary cycles. A complete analysis of these experiments and additional replicates will be presented elsewhere (Becks et al., *in prep.*)

**Systems with more than two prey types**

The details of the tradeoff between defense and competitive ability become important when more than two prey types are initially present. Depending on the shape of the tradeoff curve, it may be possible for more than two prey types to coexist with the predator. In that situation, the number of prey types present, the range of palatability between the least and most defended types, and the shape of the tradeoff curve together determine whether the system exhibits stable limit cycles, transient cycling followed by a coexistence equilibrium of one or more types, or exclusion of all but one prey genotype. The tradeoff between defense against predation and competitive ability is modeled as follows:

$$k_i(p_i) = \frac{k_C^*(1+\gamma)}{1+\gamma\, p_i^b}, \quad i = 1, 2, ..., N \tag{2}$$

where $k^*_C$ is the half-saturation constant for the completely undefended type, $\gamma$ is cost of defense, and $b$ sets the shape of the tradeoff curve. Increased defense results in an increase in the half-saturation constant $k_i(p)$, and thus a reduction in competitive ability relative to the undefended type in a predator-free environment. This curve (Fig. 4A) is only an example of many tradeoff formulations which produce consistently similar results. For our purposes here, we set the minimum p-value as $p_1 = 0.01$, and palatabilities $p_i$ take values between 0.01 and 1. Again the mathematical model for the system is equation (*1*), but in this case with $i = 1, 2, .... N$.

For shape parameter $b > 1$, the tradeoff curve is convex (Figure 4A) for most settings of the cost parameter $\gamma$ and favors extreme types (Figure 5A-C). In the limit (as time $t \to \infty$) the system behaves like a two-clone system: given $p_1$ small, and any initial number of types arrayed on the tradeoff curve, there are stable evolutionary cycles for the dilution rate $d = 0.7$ (Fig. 4, A-C). As defense drops ($p_1$ increases in value), there is first a narrow range of stable coexistence between the two extreme types, $p_1$ and $p_N$ together with the predator. As with the two-prey system, there is then a substantial range of defense levels where the steady state involves only the most defended type ($p_1$) and the predator. As the defense of the most defended type becomes moderate, like the two-prey system there are then consumer-resource cycles. The small variation in cycle period observed in Fig. 4 (B, C) for different numbers of initial prey types is a transient phenomenon, the duration of which increases as more prey types are added to the system and the tradeoff curve gets "crowded": each subsequent prey type becomes more similar to those immediately adjacent to it on the curve. Fig. 5 (A-C) shows surviving prey types for each of two initial scenarios: $N = 2$, 4, or 32 prey types when $b = 2$. In each case it is the extreme prey types (filled symbols) that survive in the presence of the predators, with the most vulnerable prey type at very low densities and the best defended at high densities.

For shape parameter $b < 1$, the tradeoff curve is concave (Fig. 4A), and favors intermediate types. Again assuming that the most defended type is nearly invulnerable to predation, $p_1 \approx 0$, as the number of prey clones increases above two types, there is an initial phase of transient evolutionary cycles followed by equilibrium during which there is a stable coexistence of two or more of the best-defended types. With an increase in prey types, the transient phase becomes shorter and shorter, and the dynamics are overall more stable with well-defended types dominating. In Fig. 4 (D-F) we illustrate what happens as prey clonal diversity is increased at the experimentally determined "cycling" range of d = 0.7, and for $p_1 \approx 0$ and tradeoff shape parameter $b = 0.5$: given $N = 2$ prey clones, the system exhibits stable evolutionary cycles with a period of about 30 days and stable cycle amplitudes. As the clones are increased to $N = 4$, the cycles rapidly shorten in period and decline in amplitude with time as the system goes to equilibrium. By $N = 32$, the transient phase comprises only one complete (evolutionary) cycle, and the system is otherwise in equilibrium. Predator densities very slowly increase with the number of initial clones, as the dominant clone shifts to an intermediate type (e.g., see Fig. 5, $b = 0.5$, $N = 32$). For further details and analysis of coexistence equilibria in our experimentally parametrized model, please see Jones & Ellner (2004).

## Results II: dynamic effects of prey and predator genetic variability

In the previous section we were able to give a rather detailed account of the effects of prey genetic diversity on community dynamics. Of course, there is no *a priori* reason why the predator in these systems should not also evolve. Indeed, we have shown that predator evolution does occur in our experimental rotifer-alga system and – just like prey evolution – can have strong effects on the stability of the observed predator-prey dynamics (Fussmann et al. 2003). We are only beginning to explore how contemporary coevolution of predator and prey affects the dynamics. Because both predator and prey in our rotifer-algal system are clonal populations, genetic diversity can easily be introduced at both trophic levels simultaneously and the effects on community dynamics be studied. Given our experience with prey-evolution-only systems we expect, however, that understanding the resulting eco-evolutionary dynamics will require a detailed mathematical analysis that accounts for the multitude of possible type-by-type interactions that occur within and across trophic levels.

We here present a first step in this direction by formulating a predator-prey model that contains up to 6 clones/genotypes within predator and prey levels and allows for the interaction of each predator with each prey type. So far, we are treating this model as an exploratory tool to understand the dynamical possibilities of this complex system and intend to model concrete coevolutionary predator-prey scenarios (i.e. rotifer and algal clones in the chemostat) in the future. The current model assumes logistic prey growth, a multispecies type-II functional response and is a system of up to 12 differential equations ($M = 6$ prey and $N = 6$ predator clones):

$$\dot{x}_i = x_i \left[ r_i \left( 1 - \sum_{i=1}^{M} x_i / K \right) - \sum_{j=1}^{N} \frac{a_{ji} y_j}{1 + H_j} \right] \, , \, i = 1,...,M$$

$$\dot{y}_j = y_j \left[ \frac{Q_j}{1 + H_j} - d_j \right] \, , \, j = 1,...,N$$

(3)

where $x_i$ and $y_i$ are the abundances of the 6 prey and predator clones, respectively and $H_j = \sum_{i=1}^{M} b_{ji} x_i, Q_j = \sum_{i=1}^{M} a_{ji} x_i$ are measures of effective total prey abundance as perceived by predator clone $j$. Parameters $r_i$ and $d_j$ are maximum prey clone growth rates and predator mortalities; parameters $a_{ji}$ and $b_{ji}$ determine the saturation value and steepness of the functional responses for pair-wise predator-prey clone interactions. The total abundance of all prey clones is regulated by the carrying capacity $K$. If no clonal diversity is present ($M = N = 1$) the system simplifies to the Rosenzweig-MacArthur model (Rosenzweig & MacArthur 1963). We parameterized in agreement with values encountered in plankton predator-prey systems (Fussmann & Heber 2002) that led to stable limit cycles for a system with a single prey and a single predator type (Fig. 6A). We introduced clonal diversity by generating $a_{ji}$ and $b_{ji}$ matrices with random uniform distribution ($\pm$ 5%) around the base values of $a = 7.5$ and $b = 5.0$. To see the effects of heritable trait variation, we contrasted the dynamics of a non-evolving (1-prey-1-predator) system using the base parameters, with the dynamics of evolving multiple-clone systems with randomly generated parameters.

What are the effects on predator-prey dynamics when diversity exists at both the predator and prey level? Numerical simulations of equation (*3*) result in a wealth of dynamic scenarios but some interesting, repeatable patterns emerge.

(1) *Selection processes among types are important and affect the dynamics at the species level*. In all our simulations only a subset of the 6 prey or predator types persisted in the long run. Selective survival of genotypes – due to natural selection – represents the evolutionary component of the eco-evolutionary dynamics we describe here. A frequently observed outcome in a 6-prey-1-predator system is the selection of a single prey type that coexists with the predator at stable equilibrium (Fig. 6B), whereas a 1-prey-1-predator system with the prey having the base parameter values displays limit cycle dynamics with the predator (Fig. 6A). Stabilization occurs because a prey type is selected that provides the highest growth rate in the presence of the predator, and at the same time shifts the predator-prey pair to the stable side of the Hopf bifurcation. Selection of genotypes at both predator and prey levels is the norm for the full 6-prey-6-predator system, but very frequently systems of multiple prey and predator types are selected that coexist on compensatory cycles (i.e. predator and prey types alternate regularly in relative frequency and display oscillatory dynamics characterized by pair-wise interactions of predator types with "their" prey type; Fig. 6 C, E). Fig. 6E shows an interesting example of coexistence of 3 predator and 3 prey types. Each of the 3 distinguishable predator-prey pairs shows characteristic predator-prey dynamics with the typical shift between predator and prey peaks. One predator-prey pair cycles much more slowly than the other two, whose frequencies are about 5 times higher. These examples show that when prey and predator are evolving in tandem the "cryptic cycles" phenomenon can occur at both trophic levels, with rapid dynamics at the clone-frequency level resulting in total species abundances that remain nearly constant.

(2) *High type diversity is rare.* Most frequently systems with two predator and two prey clones become selected. Complicated systems with many types appear to be unable to coexist in the homogeneous environment that this model represents. On the other hand, selection of just one clone of prey and predator each is not the most frequent outcome. It seems that our 6-by-6 interaction model is able to reproduce the essential dynamical patterns expected in homogeneous systems of much higher genetic diversity because they all reduce to systems of low to moderate diversity. Clonal genetic diversity in natural communities is often much higher (e.g. De Meester et al. 2006). This suggests that complex dynamics in complex community networks are not a likely mechanism that maintains high levels of genetic diversity in homogenous natural systems.

(3) *Genotype dynamics are wild, species and community dynamics are calm.* The increased complexity of the multi-predator-prey system tends to lead to more complex and erratic dynamics at the genotype level (Fig. 6C, E, G), compared with the benchmark one-predator-one-prey system (Fig. 6A). However, species dynamics (the sums of the abundances of predator and prey types, respectively) become stabilized relative to the benchmark system, as also observed by Vos et al. (2001) for a similar model system. This is partly due to statistical averaging of time series (Cottingham et al. 2001), but is primarily caused by the compensatory nature of the dynamics, i.e. the regular succession of different clones as opposed to their synchronous appearance and disappearance, and may lead to apparently steady-state dynamics at the species level (Fig. 6D, F). This is another instance of cryptic dynamics in which highly dynamic patterns of cyclical selection of genotypes are hidden under a calm surface of species dynamics. Because this type of cryptic dynamics can occur at very moderate levels of trait diversity (parameters *a* and *b* diverging only up to 5% from their mean value) it is likely to play an important role in real communities, even when within-species diversity is relatively low.

(4) *Species dynamics may appear noisy or display intermittency.* Due to the complexity of the dynamics, genotype abundances never add up to smooth steady state or cyclical species dynamics. Instead, equilibria appear "noisy" (Fig. 6D, F) or the dynamics show intermittent patterns. Intermittency occurs because periods dominated by multiple genotype interactions alternate with periods of classical predator-prey oscillations (when only one prey and one predator genotype dominate; Fig. 6G, H). These simulations suggest that the noisy and intermittent patterns so frequently observed in natural communities are potentially due to intrinsic, cryptic genotype dynamics and are not necessarily the result of changing environmental factors.

The patterns described here emerged from a relatively simple mathematical model in which we – somewhat artificially – generated the potential for evolutionary dynamics by setting initial levels of genetic diversity of up to 12 genotypes. As such, we investigated the interaction between ecological dynamics and the dynamical process of natural selection, but neglected processes that are able to restore genetic diversity. In natural communities, mutation and immigration of genotypes are such processes and future studies should consider their impact on eco-evolutionary dynamics. We suspect that the effects of pulsed mutation or immigration events will be understandable within the framework we presented here because they just reset genetic diversity to higher levels at fixed time intervals (as we did at the start of our simulations). However, continuous rates of mutation or immigration at the time scale of the ecological dynamics may lead to dynamical outcomes not covered by our current analysis and we see the inclusion of these processes as an important continuation of our work.

## Discussion

Although the fields of ecology and evolutionary biology are both over a century old, we are only beginning to understand how tightly coupled the processes at the heart of these disciplines truly are. Becoming somebody else's dinner represents very strong natural selection, and it is hard to imagine the predator-prey interaction or any other strong interspecific interaction occurring without direct evolutionary consequences. But exactly that thinking is embedded in much of the ecological theory underpinning environmental and natural resource management. By studying some simple model food webs based directly on experiments, we hope to start developing ecological theory for a rapidly evolving biosphere.

We return now to the questions we posed in the Introduction: how rapid contemporary evolution can affect the dynamics of communities (Question 1) and then comparing the dynamic possibilities when genetic diversity is present versus absent (Question 2). Even our simplest experimental systems (one predator, two prey genotypes) and the models developed to explain them generate a range of qualitatively different dynamics (Results I). Changing the details of a tradeoff between the cost of a defense, and its effectiveness can dramatically alter the identity of the surviving prey types, as can environmental factors (such as microcosm flow-through rate). Adding realistic, if theoretical, levels of evolutionary complexity in the form of heritable variation in ecologically relevant traits of both predator and prey opens the door to even more possibilities (Results II): selection of one prey type under some circumstances, and multiple types under others, and dynamics varying from equilibrium to chaos. Conversely, evolutionary dynamics can sometimes make for greater simplicity at the species or community level, as complex dynamics at the genotype level may lead to constant or nearly constant population abundances, instead of cycles, in coexisting populations (Results I and II).

A challenge posed by these findings is to develop general theory for linked eco-evolutionary dynamics, so that we can understand which biological conditions give rise to different ecological outcomes, and why. One approach that we are currently exploring is the theory of slow-fast systems. This was used a decade ago (Khibnik and Kondrashov 1997) under the conventional assumption of slow evolutionary dynamics relative to fast ecological dynamics. To gain insight into possible effects of coupled and equally rapid evolutionary and ecological dynamics, we believe that it will be useful to consider the opposite limit where evolutionary change is fast relative to ecological dynamics (Cortez and Ellner, *in prep*). This is not biologically impossible – it could happen if many births, balanced by nearly as many deaths, result in slow changes in total population size but allow rapid change in genotype frequencies. Most importantly, it reduces model dimension and creates real possibilities for mapping out all dynamic possibilities and understanding when each of them can occur.

Because temporal dynamics provide information about underlying processes (e.g., Kendall et al. 2005, Ives et al. 2008), large-scale changes such as consumer-resource cycles are especially revealing about potential effects of rapid contemporary evolution. Several types of natural large-scale ecological dynamics have offered opportunities for studying rapid evolution in the wild, including invasions (e.g., Lavergne & Molofsky 2007, Kinnison et al. 2008), range expansion (Reznick et al. 2008), periodic outbreaks of infectious diseases (e.g., Elderd et al. 2008), annual population cycles driven by seasonality (Duffy & Sivars-Becker 2007), and trait cycles (Sinervo & Lively 1996, Sinervo et al. 2000). Systems at steady state may also may be shaped by rapid evolution, for example population stability may be an outcome of the interaction between potentially rapid ecological and evolutionary processes (e.g., Doebeli & Koella 1995; Zeineddine & Jansen 2005). But demonstrating what drives this outcome in natural settings may require experimental manipulations that are necessarily high-cost and high-effort (e.g., Fischer et al. 2001), and whose interpretation has often been problematic because of inadvertent confounding factors resulting from large-scale interventions (Turchin 2003). Natural dynamic processes may thus offer the best prospects for confronting theory with data on consequences of rapid evolution, which is critically important as the theory begins to develop.

Our most fundamental theoretical prediction is the level of heritable variation in a trait affecting ecological interactions is a bifurcation parameter: small gradual quantitative changes in unseen evolutionary processes can cause large abrupt qualitative changes at the population and community levels. For the practical issues of ecological management and forecasting that we posed in the Introduction, it is already widely recognized that we need to consider possible effects of evolutionary responses that are evoked by human impacts and interventions. Our results raise the converse issue: we may also need to consider the effects of evolutionary responses that are slowed or constrained when genetic variation is reduced by human impacts such as habitat loss or fragmentation.

**Acknowledgements.** Our research has been supported primarily by the Andrew W. Mellon Foundation, the James S. McDonnell Foundation, and the Natural Sciences and Engineering Council of Canada. We thank Andrew Hendry and two anonymous referees for their comments on the manuscript, and also the many Cornell undergraduates who assisted with data collection.

# References

Abrams, P. A. 2005 'Adaptive Dynamics' vs. 'adaptive dynamics'. Journal of Evolutionary Biology 18, 1162-1165.

Abrams, P. A. & Matsuda, H. 1997 Prey adaptation as a cause of predator-prey cycles. *Evolution* 51, 1742-1750.

Andersson, D. L. & Levin, B. R. 1999 The biological cost of antibiotic resistance. *Current Opinion in Microbiology* 2, 489–493.

Boraas, M. E., Searle, D. B. & Boxhorn, J. E. 1998 Phagotrophy by a flagellate selects for colonial prey: A possible origin of multicellularity. *Evolutionary Ecology* 12, 153-164.

Cain, M. L., Bowman, W. D. & Hacker, S. D. 2008 *Ecology*. Sinauer Associates, Sunderland, MA. 621 pp.

Cottingham, K. L., Brown, B.L. & Lennon, J. T. 2001 Biodiversity may regulate the temporal variability of ecological systems. *Ecology Letters* 4, 72-85.

Cousyn, C., De Meester, L., Colbourne, J. K., Brendonck, L., Verschuren, D. & Volckaert, F. 2001 Rapid, local adaptation of zooplankton behavior to changes in predation pressure in the absence of neutral genetic changes. *Proceedings of the National Academy of Sciences of the United States of America* 98, 6256–6260.

De Meester, L., Vanoverbeke, J., De Gelas, K., Ortells, R. and Spaak, P. 2006 Genetic structure of cyclic parthenogenetic zooplankton populations – a conceptual framework. *Archiv für Hydrobiologie* 167, 217-244.

Doebeli, M. & Koella, J. C. 1995 Evolution of simple population-dynamics. *Proceedings of the Royal Society of London Series B-Biological Sciences* 260**,** 119-125.

Dube, D., Kim, K., Alker, A. P. and Harvell, C. D. 2002 Size structure and geographic variation in chemical resistance of sea fan corals *Gorgonia ventalina* to a fungal pathogen. *Mar. Ecol. Prog. Ser.* 231, 139-150.

Duffy, M. A. & Sivars-Becker, L. 2007 Rapid evolution and ecological host-parasite dynamics. *Ecology Letters* 10, 44-53.

Decaestecker, E., Gaba, S., Raeymaekers, J. A. M., Stoks, R., Van Kerckhoven, L., Ebert, D. & De Meester, L. 2007 Host-parasite 'Red Queen' dynamics archived in pond sediment. *Nature* 450, 870-873.

Eldred, B. D., Dushoff, J. & Dwyer, G. 2008 Host-pathogen interactions, insect outbreaks, and natural selection for disease resistance. *American Naturalist* 172, 829-842.

Ellner, S., Hairston, N. G. Jr., Kearns, C. M. & Babaï, D. 1999 The roles of fluctuating selection and long-term diapause in microevolution of diapause timing in a freshwater copepod. *Evolution* 53, 111-122.

Ezard, T. H. G., Clutton-Brook, T. H., Côté, S. D. & Pelletier, F. (This Volume).

Ferrière, R. & Couvert, D. 2004 *Evolutionary conservation biology*, vol. 4 of Cambridge Studies in adaptive dynamics. Cambridge University Press, Cambridge.

Fischer, J. M., Klug, J. L., Ives, A. R. & Frost, T. M. 2001 Ecological history affects zooplankton community responses to acidification. *Ecology* 82, 2984-3000.

Fussmann, G. F., Ellner, S. P. & Hairston, N. G. Jr. Evolution as a critical component of plankton dynamics. *Proceedings of the Royal Academy of London B Biological Sciences* 270, 1015-1022.

Fussmann, G. F., Ellner, S. P., Shertzer, K. W. & Hairston, N. G. Jr. 2000 Crossing the Hopf bifurcation in a live predator-prey system. *Science* 290, 1358–1360.

Fussmann, G. F. & Heber, G. 2002 Food web complexity and chaotic population dynamics. *Ecology Letters* 5, 394-401.

Fussmann, G. F., Loreau, M. & Abrams, P. A. 2007 Eco-evolutionary dynamics of communities and ecosystems. *Functional Ecology* 21, 465-477.

Gagneux, S., Long, C. D., Small, P. M., Van, T., Schoolnik, G. K. & Bohannan, B. J. M. 2006 The competitive cost of antibiotic resistance in *Mycobacterium tuberculosis*. *Science* 312, 1944–1946.

Grant, P. R. & Grant, B. R. 2002 Unpredictable evolution in a 30-year study of Darwin's finches. *Science* 296, 707–711.

Hairston, N. G. Jr. & Walton, W. E. 1986 Rapid evolution of a life-history trait. *Proceedings of the National Academy of Sciences USA* 83, 4831-4833.

Hairston, N. G. Jr. & De Meester, L. 2008 *Daphnia* paleogenetics and environmental change: deconstructing the evolution of plasticity. *International Review of Hydrobiology* 93, 578-592.

Hairston, N. G. Jr., Ellner, S. P., Geber, M. A., Yoshida, T. & Fox, J. A. 2005 Rapid evolution and the convergence of ecological and evolutionary time. *Ecology Letters* 8, 1114–1127.

Hairston, N. G. Jr., Lampert, W., Caceres, C. E., Holtmeier, C. L., Weider, L. J., Gaedke, U., Fischer, J. M., Fox, J. A. & Post, D. M. 1999 Lake ecosystems – rapid evolution revealed by dormant eggs. *Nature* 401, 446–446.

Heath, D. D., Heath, J. W., Bryden, C. A., Johnson, R. M. & Fox, C. W. 2003 Rapid evolution of egg size in captive salmon. *Science* 299, 1738–1740.

Hendry, A. P., Farrugia, T. & Kinnison, M. T. 2008 Human influences on rates of phenotypic change in wild animal populations. *Molecular Ecology* 17, 20-29.

Hendry, A. P. & Kinnison, M. T. 1999 The pace of modern life: measuring rates of contemporary microevolution. *Evolution* 53, 1637-1653.

Ives, A. R., Einarsson, Á., Jansen, V. A. A. & Gardarsson, A. 2008 High-amplitude fluctuations and alternative dynamical states of midges in Lake Myvatn. Nature 452, 84-87.

Johnson, M. T. J., Lajeunesse, M. J. & Agrawal, A. 2006 Additive and interactive effects of plant genotypic diversity on arthropod communities and plant fitness. *Ecology Letters* 9, 24-34.

Johnson, M. T. J., Vellend, M. & Stinchcombe, J. R. (This Volume).

Jones, L. E. & Ellner, S. P. 2004 Evolutionary tradeoff and equilibrium in an aquatic predator-prey system. *Bulletin of Mathematical Biology* 66, 1547–73.

Jones, L. E. & Ellner, S. P. 2007 Effects of rapid prey evolution on predator-prey cycles. *Journal of Mathematical Biology* 55, 541–573.

Kendall, B. E., Ellner, S.P., McCauley, E., Wood, S. N., Briggs, C. J., Murdoch, W. W. & Turchin, P. 2005 Population cycles in the pine looper moth *Bupalus piniarius*: dynamical tests of mechanistic hypotheses. *Ecological Monographs* 75, 259–276

Khibnik, A. I. & Kondrashov, A. S. 1997 Three mechanisms of Red Queen dynamics. *Proceedings of the Royal Society of London Series B-Biological Sciences*, 264, 1049-1056

Kinnison, M. T & Hendry, A. P. 2001 The pace of modern life II: from rates of contemporary microevolution to pattern and process. *Genetica* 112-113, 45-164.

Kinnison, M. T & Hairston, N. G. Jr. 2007. Eco-evolutionary conservation biology: contemporary evolution and the dynamics of persistence. *Functional Ecology*, 21 444–454.

Kinnison, M. T., Unwin, M. J. & Quinn, T. P. 2008 Eco-evolutionary vs. habitat contributions to invasion in salmon: experimental evaluation in the wild. *Molecular Ecology* 17, 405-414.

Lavergne, S. & Molofsky, J. 2007 Increased genetic variation and evolutionary potential drive the success of an invasive grass. *Proceedings of the National Academy of Sciences USA* 104, 3883-3888.

Lenski, R. E. & Travisano, M. 1994 Dynamics of adaptation and diversification: a 10,000-generation experiment with bacterial populations. *Proceedings of the National Academy of Sciences USA* 91, 6808-6814**.**

Meyer, J. R., Ellner, S. P., Hairston, N. G. Jr, Jones, L. E. & Yoshida, T. 2006 Prey evolution on the time scale of predator-prey dynamics revealed by allele-specific quantitative PCR. *Proceedings of the National Academy of Sciences of the United States of America* 103, 10690–10695.

Meyer, J. R. & Kassen, R. 2007 The effects of competition and predation on diversification in the model adaptive radiation. *Nature* 446, 432-435.

Molles, M. C. Jr. 2008 *Ecology: concepts, applications*. 4th Ed., McGraw Hill, New York, NY, USA.

Olsen, E. M., Heino, M., Lilly, G. R., Morgan, M. J., Brattey, J., Ernande, B.& Dieckmann, U. 2004 Maturation trends indicative of rapid evolution preceded the collapse of northern cod. *Nature* 428, 932–935.

Pelletier, F., Clutton-Brock, T., Pemberton, J., Tuljapurkar, S. & Coulson, T. 2007 The evolutionary demography of ecological change: linking trait variation and population growth. *Science* 315, 1571-1574.

Palkovacs, E. P., Marshall, M. C., Lamphere, B. A., Lynch, B. R., Weese, D. J., Fraser, D. F., Reznick, D. N., Pringle, C. M. & Kinnison, M. T. (This Volume).

Palumbi, S. R. 2001 *The evolution explosion: how humans cause rapid evolutionary change*. Norton, New York, NY. 277 pp.

Post, D. & Palkovacs, E. P. (This Volume).

Reznick, D. N. & Ghalambor, C. K. 2001 The population ecology of contemporary adaptations: what empirical studies reveal about the conditions that promote adaptive evolution. *Genetica* 112, 183–198.

Reznick, D. N., Ghalambor, C. K. & Crooks, K. 2008 Experimental studies of evolution in guppies: a model for understanding the evolutionary consequences of predator removal in natural communities. *Molecular Ecology* 17, 97-107.

Reznick, D. N., Shaw, F. H., Rodd, F. H. & Shaw, R. G. 1997 Evaluation of the rate of evolution in natural populations of guppies (*Poecilia reticulata*). *Science* 275, 1934–1937.

Rosenzweig, M. L. & MacArthur, R.H. 1963 Graphical presentation and stability conditions of predator-prey interactions. *American Naturalist* 97, 209-223.

Shertzer, K. W., Ellner, S. P., Fussmann, G. F. & Hairston, N. G. Jr. 2002 Predator-prey cycles in an aquatic microcosm: testing hypotheses of mechanism. *Journal of Animal Ecology* 71, 802–815.

Sinervo, B. & Lively, C. M. 1996.  The rock-paper-scissors game and the evolution of alternative male strategies. *Nature* 380, 240-243.

Sinervo, B., Svensson, E. & Comendant, T.,  2000.  Density cycles and an offspring quantity and quality game driven by natural selection. *Nature* 406, 985–988.

Strauss, S.Y., Lau, J. A., Schoener, T.W. & Tiffin, P. 2008 Evolution in ecological field experiments: implications for effect size. *Ecology Letters* 11, 199-207.

Swain, D. P., Sinclair, A. F. & Mark Hanson, J. , 2007.  Evolutionary response to size-selective mortality in an exploited fish population. *Proceedings of the Royal Society of London Series B-Biological Sciences* 274, 1015–1022.

Thompson, J. N. 1998 Rapid evolution as an ecological process. *Trends in Ecology and Evolution* 13, 329–332.

Turchin, P. 2003 *Complex Population Dynamics*. Princeton Univ. Press, Princeton, NJ.

Urban, M. C., Leibold, M. A., Amarasekare, P., De Meester, L., Gomulkiewicz, R., Hochberg, M. E., Klausmeier, C. A., Loeuille, N., de Mazancourt, C., Norberg, J., Pantel, J. H., Strauss, S. Y., Vellend, M. & Wade, M. J. 2008 The evolutionary ecology of metacommunities. *Trends in Ecology and Evolution* 23, 311-317.

Vos, M., Moreno Berrocal, S., Karamaouna, F., Hemerick, L. & Vet, L. E. M. 2001 Plant-mediated indirect effects and the persistence of parasitoid-herbivore communities. *Ecology Letters* 4, 38-45.

Whitham, T. G., Bailey, J. K., Schweitzer, J. A., Shuster, S. M., Bangert, R. K., Leroy, C. J., Lonsdorf, E. V., Allen, G. J., DiFazio, S.P., Potts, B. M., Fischer, D. G., Gehring, C. A., Lindroth, R. L., Marks, J. C., Hart, S. C., Wimp, G. M. & Wooley, S. C. 2006 A framework for community and ecosystem genetics: from genes to ecosystems. *Nature Reviews Genetics* 7, 510-523.

Yoshida, T., Ellner, S. P., Jones, L. E., Bohannan, B. J. M., Lenski, R. E. & Hairston, N. G. Jr. 2007 Cryptic population dynamics: Rapid evolution masks trophic interactions. *PloS Biology* 5, 1868–1879.

Yoshida, T., Hairston, N. G. Jr. & Ellner, S. P. 2004 Evolutionary trade-off between defence against grazing and competitive ability in a simple unicellular alga, *Chlorella vulgaris. Proceedings of the Royal Society of London Series B-Biological Sciences* 271, 1947–1953.

Yoshida, T., Jones, L. E., Ellner, S. P., Fussmann, G. F. & Hairston, N.G. Jr. 2003 Rapid evolution drives ecological dynamics in a predator-prey system. *Nature* 424, 303–306.

Zeineddine, M. & Jansen, V. A. A. 2005 The evolution of stability in a competitive system. *Journal of Theoretical Biology* 236**,** 208-215.

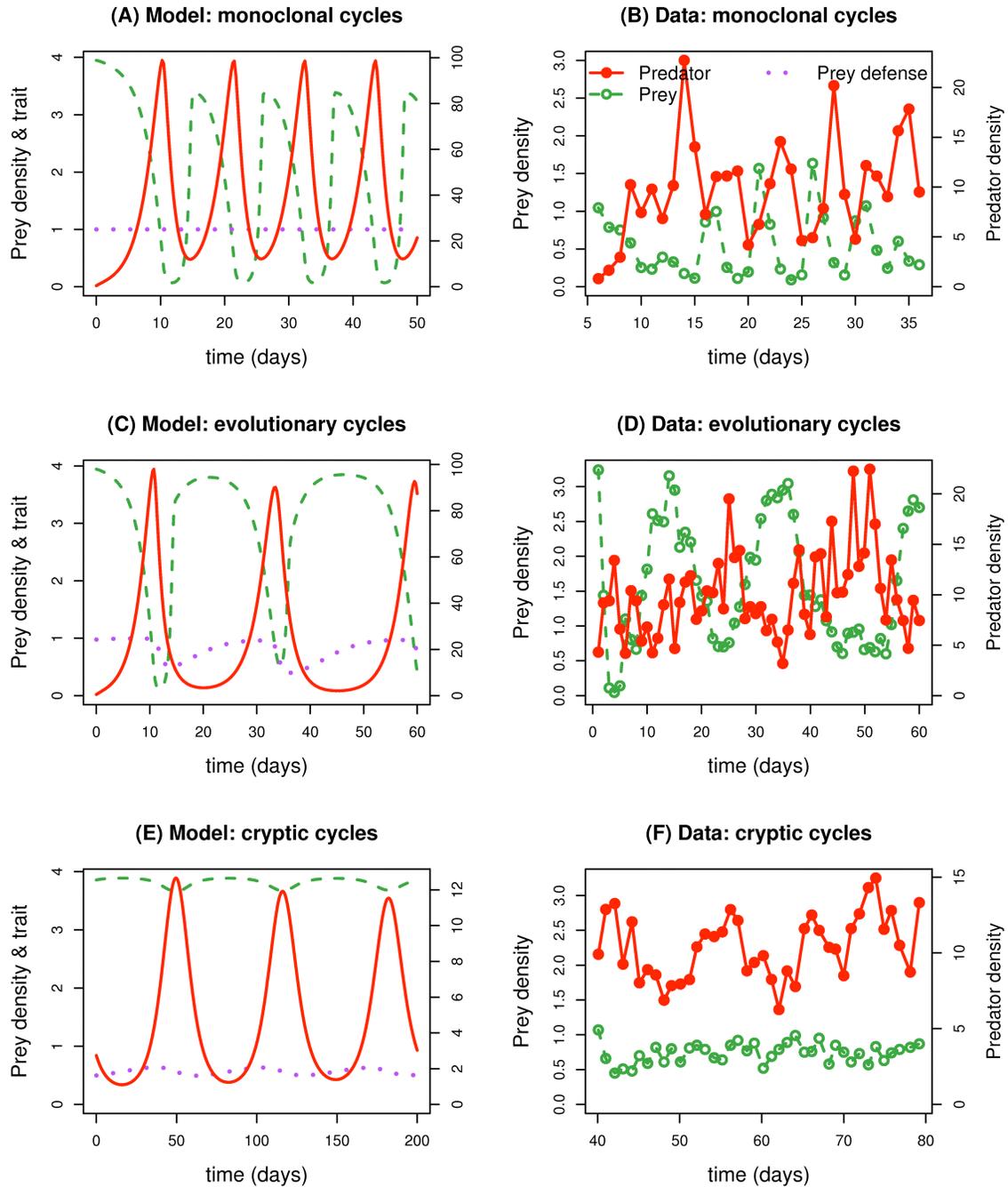

**Figure 1:** Comparison of theoretical and experimental results for rotifer-alga chemostats with the rotifer *Brachionus calyciflorus* as the predator and the asexual green alga *Chlorella vulgaris* as the prey. (A, B): classical predator-prey cycles when algae are monoclonal (one genotype) and so cannot evolve (red solid curve is the predator, green dashed curve is the prey, purple dotted curve is the average prey vulnerability to predation with 1=no defense, 0=complete invulnerability to attack). Data from Yoshida et al. (2003). (C, D): evolutionary cycles when prey are genetically variable. Data from Fussmann et al. (2000). (E, F): cryptic cycles when prey are genetically variable with low cost of defense. Data from Yoshida et al. (2007); spectral analysis confirms the presence of periodicity in the predator dynamics, $P < 0.01$. Units: $10^6$ cells/ml (algae), females/ml (rotifers), mean palatability (prey trait). Additional experimental replicates for panels B,D, and F can be found in the publications cited as the source for the data.

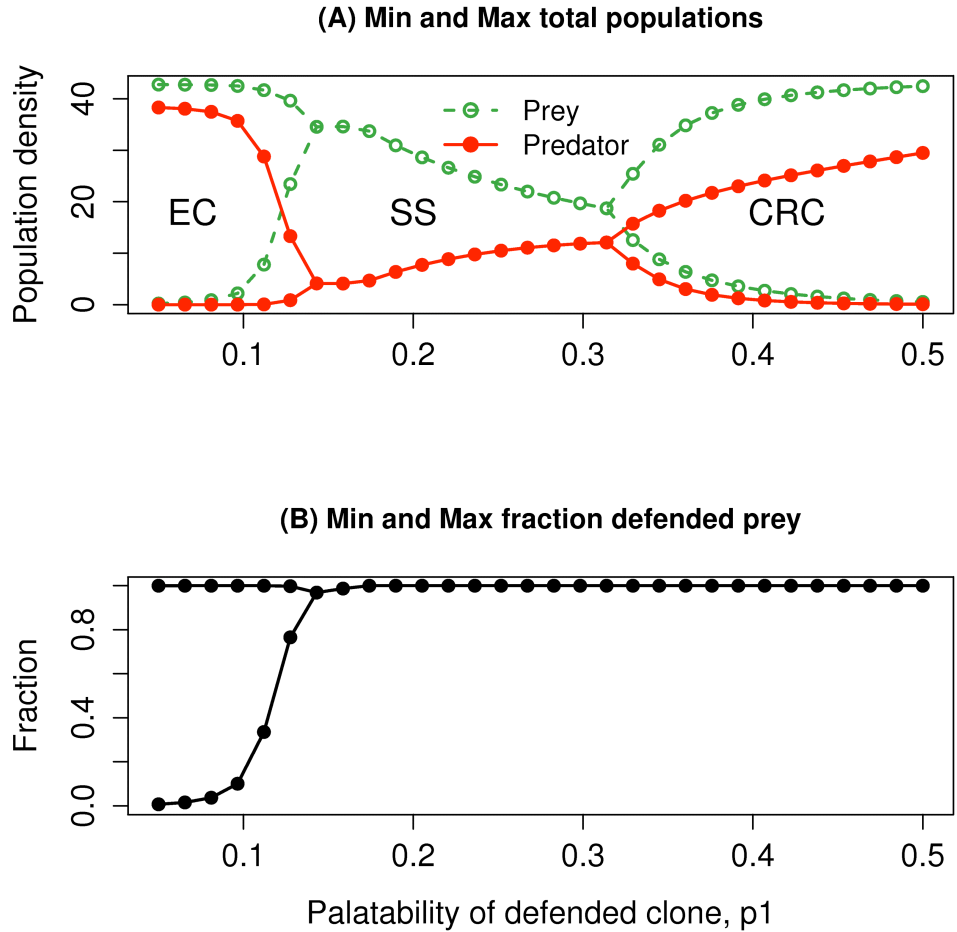

**Figure 2:** Bifurcation diagram for a model with two prey clones having palatabilities $p_1$, and $p_2 = 1$, with a linear tradeoff between $p$ and competitive ability. The top plot shows the minimum and maximum of prey (green) and predator (red) populations, once the populations have settled onto their attractor (an equilibrium or limit cycle). EC=evolutionary cycles, SS=steady state, CRC=classical consumer-resource cycles. The bottom plot shows the minimum and maximum of the fraction of the defended clone, $p_1$, in the prey population.

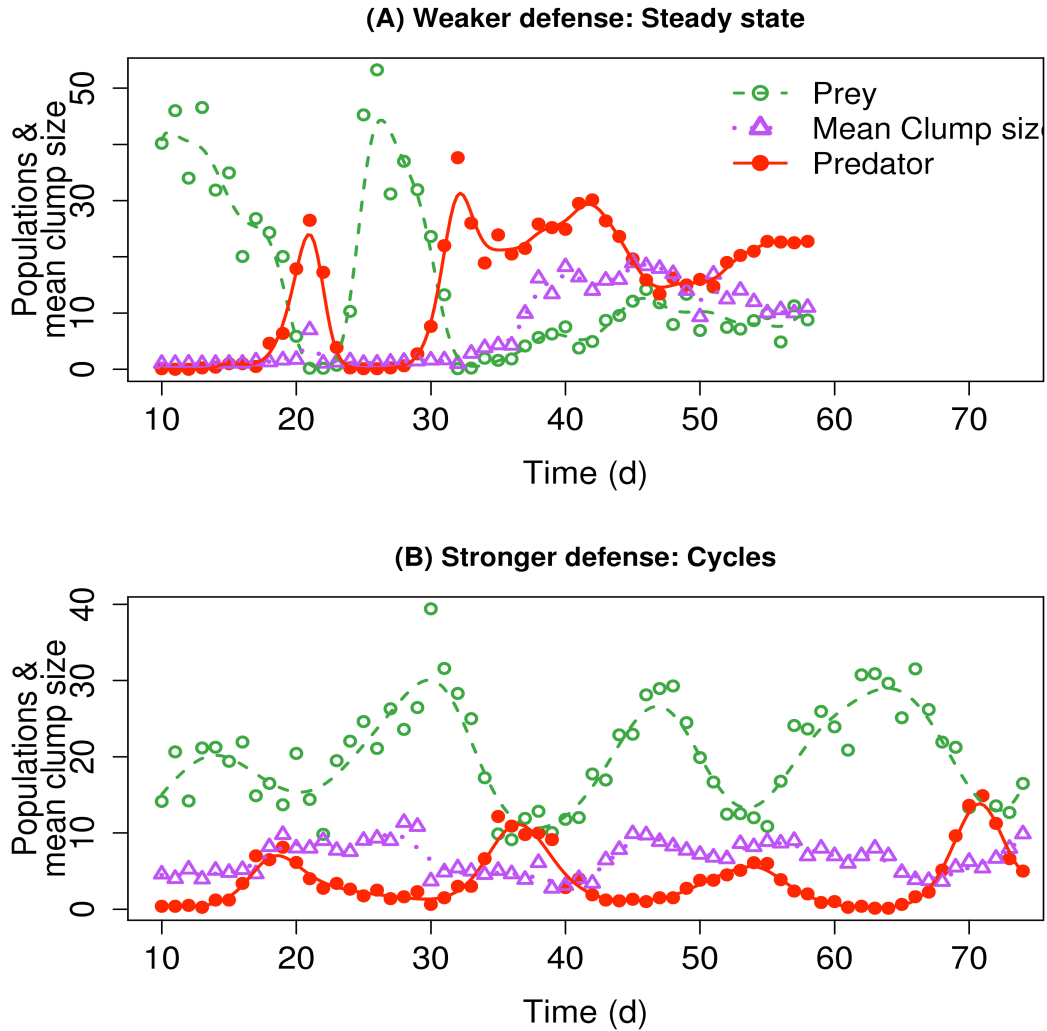

**Figure 3:** Chemostat experiments (L. Becks, unpublished data) with the rotifer *Brachionus calyciflorus* as the predator (red solid curve and filled circles) and the asexual green alga *Chlamydomonas reinhardii* as the prey (green dashed curve and open circles). The symbols are experimental measurements, and the curves are a smooth of the data by local polynomial regression. The purple curve (open triangles, dotted curve) shows the mean clump size of the algal population. See the text for discussion of clumping as an anti-predator defense trait. Units are individuals/ml for predator population density, $10^4$ cells/ml for prey population density, and mean number of cells per clump.

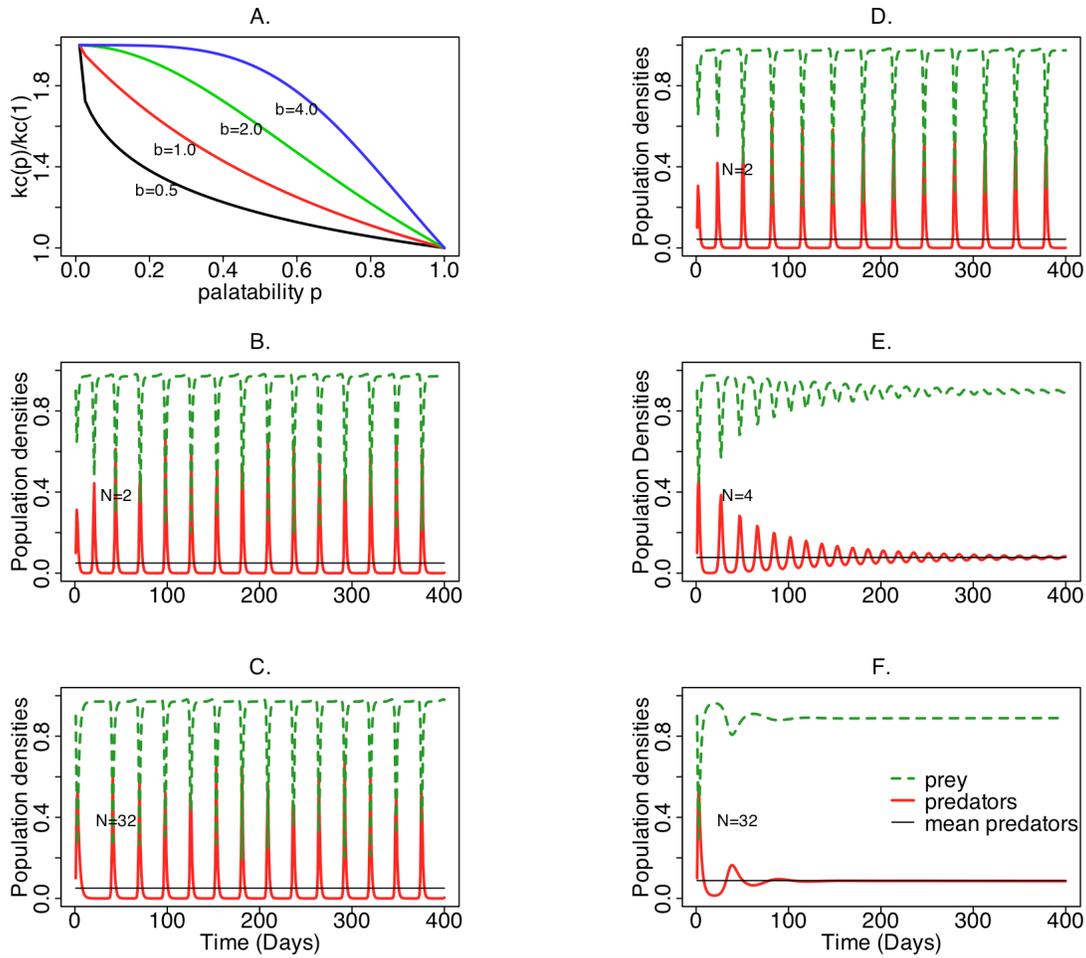

**Figure 4**: (A): tradeoff curves specified by equation (2). (B, C): trajectories for a tradeoff curve with $b = 2$, $\gamma = 1$, for $N = 2$ (B), and 32 (C) initial prey types. Total prey are shown in dashed green and the predator is shown in solid red line. A thin black line denotes the mean predator density, which remains approximately 0.05 for all three scenarios shown on the left. The dilution rate is $d' = 0.7(\mathrm{d}^{-1})$, a setting for which we consistently observed cycles in our original experimental system (Jones & Ellner 2004). (D-F): trajectories for a tradeoff curve with $b = 0.5$, $\gamma = 1$, for $N = 2$ (D), 4 (E) and 32 (F) initial prey types. Total prey are shown in green and the predator is shown in red line. A thin black line denotes the mean predator density, which is approximately 0.05, 0.07, and 0.09, respectively, for the 2, 4, and 32 clone scenarios shown on the right.

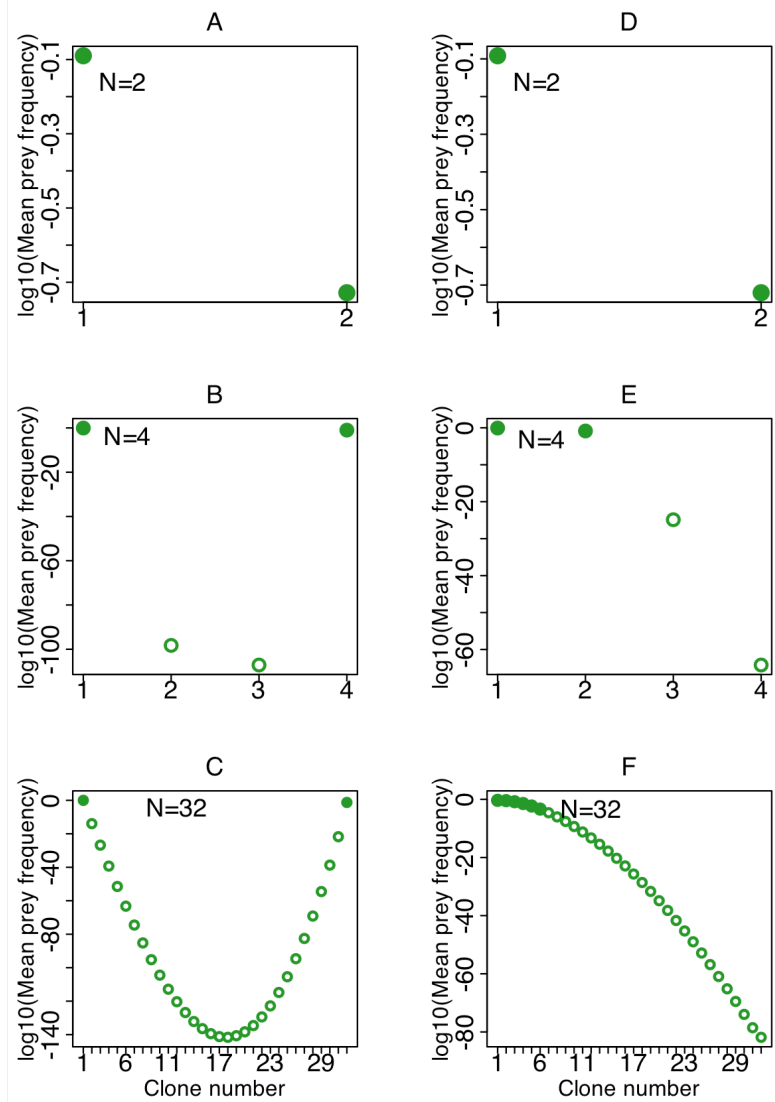

**Figure 5:** Mean clonal frequencies (log10 scale, note differences in vertical scaling) for surviving prey types after 600 days at a dilution rate $d' = 0.7(d^{-1})$. (A-C), assuming that there were initially $N = 2$, 4 and 32 prey types arrayed on a tradeoff curve with shape parameter $b = 2$ and cost parameter $\gamma = 1$. In this case, the surviving clones are the extremes (filled circles), and after the transient phase - which increases in length as prey types are added to the tradeoff curve - the system behaves approximately like a two clone system: it exhibits stable limit cycles at this dilution rate. (D-F), assuming that there were initially $N = 2$, 4 and 32 prey types arrayed on a tradeoff curve with shape parameter $b = 0.5$ and cost parameter $\gamma = 1$. In this case there is an internal equilibrium with the most defended type and one or more very well defended types (filled circles). The number of coexisting types increases with the number of clones and the concavity of the tradeoff curve. Clones are indexed in the order of increasing palatability.

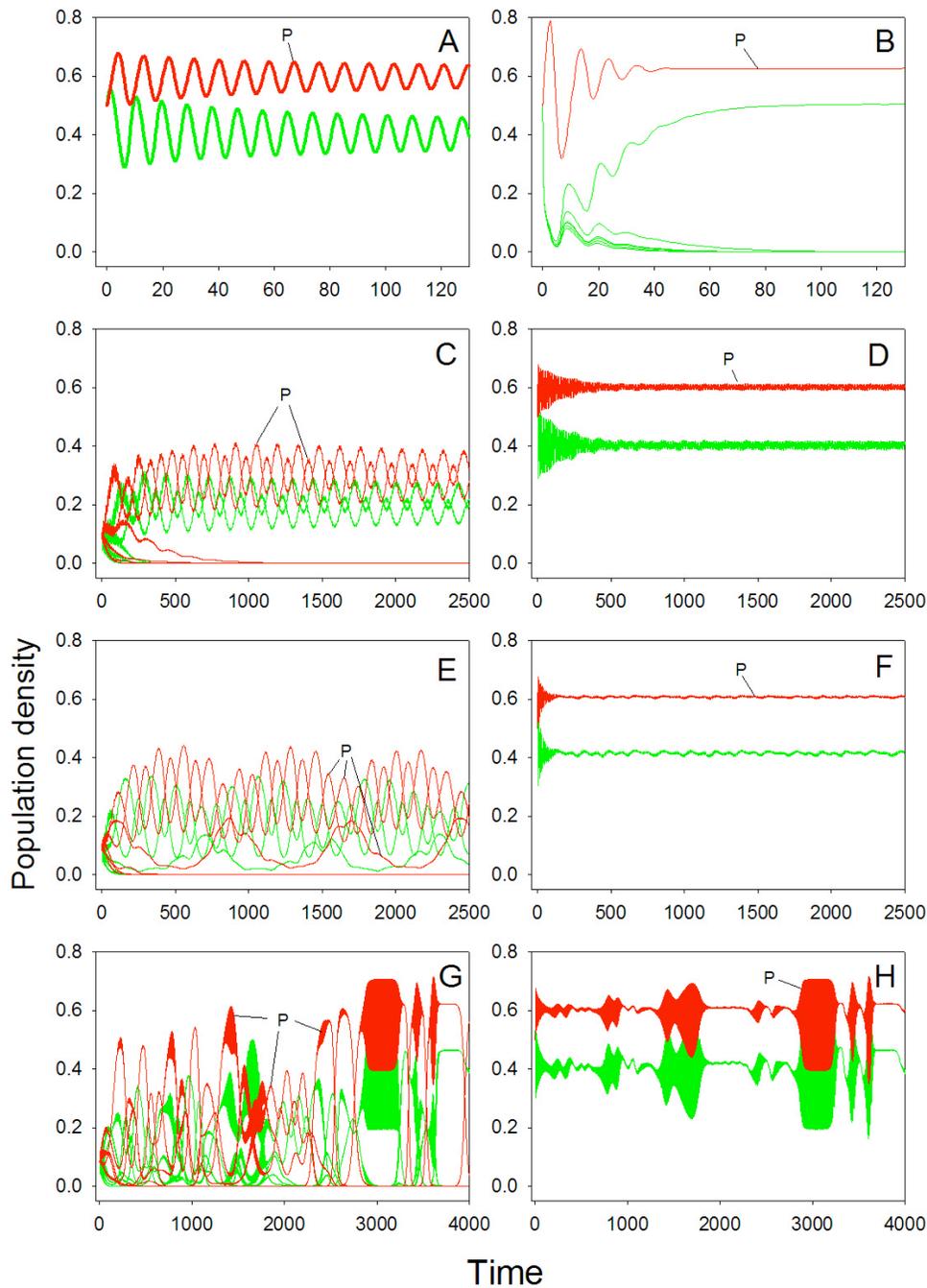

**Figure 6:** Trajectories of predator-prey systems (equation (*3*)) with multiple prey (green) and predator (red; letter "P") genotypes. (A) Benchmark system with one predator and one prey genotype. (B) Selection of a single prey type in 6-prey-1-predator system. (C) Selection of a 2-prey-2-predator system in a 6-prey-6-predator system. (D) Species dynamics of system C (summation of predator and prey genotype abundances). (E) Selection of a 3-prey-3-predator system in a 6-prey-6-predator system. Note the two fast and one slowly cycling predator-prey pairs. (F) Species dynamics of system E. (G) Intermittent dynamics alternating between periods of dominance of single and multiple predator-prey pairs. (H) Species dynamics of system G. Parameters: $r_i = 2.5$, $K = 1$, $d_j = 1$, $a_{ji} = 7.5 \pm 5\%$, $b_{ji} = 5.0 \pm 5\%$. High-frequency oscillations appear as continuously filled areas in D, F, G, H.